\documentclass[twocolumn,prb,showpacs]{revtex4}
\usepackage{graphicx}
\usepackage{dcolumn}
\usepackage{bm}
\usepackage{color}

\begin{document}
\title{Coexistence of distinct charge fluctuations in $\theta$-(BEDT-TTF)$_2$X}
\author{S.~Nishimoto,$^{1,*}$ M.~Shingai,$^2$ and Y.~Ohta$^2$}
\affiliation{$^1$Max-Planck-Institut f\"ur Physik komplexer Systeme, 
D-01187 Dresden, Germany\\
$^2$Department of Physics, Chiba University, Chiba 263-8522, Japan}
\date{\today}
\begin{abstract}
Using the Lanczos exact-diagonalization and density-matrix renormalization 
group methods, we study the extended Hubbard model at quarter filling defined 
on the anisotropic triangular lattice.  We focus on charge ordering (CO) 
phenomena induced by onsite and intersite Coulomb interactions.  
We determine the ground-state phase diagram including three CO phases, 
i.e., diagonal, vertical, and three-fold CO phases, based on the calculated 
results of the hole density and double occupancy.  We also calculate the 
dynamical density-density correlation functions and find possible 
coexistence of the diagonal and three-fold charge fluctuations in a certain 
parameter region where the onsite and intersite interactions compete.  
Furthermore, the characteristic features of the optical conductivity 
for each CO phase are discussed.  
\end{abstract}
\maketitle

\section{Introduction}

In a number of correlated electron systems, ordering of charge degrees of 
freedom of electrons, or charge ordering (CO), is a key feature to 
understanding their physical properties because it is intimately related 
to the electronic transport and magnetic phenomena.~\cite{Tokura00}  
The concept of CO was originally introduced to interpret an array order 
of Fe$^{2+}$ and Fe$^{3+}$ ions in magnetite Fe$_3$O$_4$, which is known 
as the Verwey transition.~\cite{Verwey47}  For several years, the importance 
of CO has been increasingly recognized in conjunction with spatial localization 
of charge carriers observed in manganese oxide La$_{1-x}$Ca$_x$MnO$_3$ 
($x \ge 0.5$),~\cite{Mori98-1} copper oxide La$_{1.6-x}$Nd$_{0.4}$Sr$_x$CuO$_4$ 
($x=0.12$),~\cite{Tranquada95} etc.  Especially, in the field of 
low-dimensional organic materials, the CO itself has been one of the main 
issues for discussion.~\cite{Lang03}  Quite recently, peculiar charge 
fluctuations have been found in the CO state of quasi-two-dimensional (2D) 
organic conductors $\theta$-(BEDT-TTF)$_2$X (hereafter, BEDT-TTF is 
abbreviated as ET) and they have attracted much attention due to their 
intriguing relation to the superconducting (SC) state.

The crystal structure of $\theta$-(ET)$_2$X are built with an alternating 
stack of ET conducting and anion (X) insulating layers.  In the conducting 
layer, the ET molecules form an anisotropic triangular-lattice structure 
(see Fig.~\ref{lattice}) and the band is $\frac{3}{4}$-filled in terms of 
electrons or $\frac{1}{4}$-filled in terms of holes.  
In experiments, the transfer integrals ($t_p, t_c$) are controllable by 
altering a dihedral angle ($\theta$) with the substitution of anion and/or 
adjustment of pressure.~\cite{Mori98,Mori00}  A decrease in the dihedral 
angle corresponds to an increase in $t_p$ and therefore a decrease in 
$t_c/t_p$.  At low temperatures ($T \lesssim 250$ K), the compounds exhibit 
a variety of phases, such as CO insulator, paramagnetic metal, and SC, 
as a function of the dihedral angle.  Overall, the metal-insulator 
transition temperature ($T_{\rm CO}$) decreases with decreasing the dihedral 
angle.  Only $\theta$-(ET)$_2$I$_3$ that has the smallest dihedral angle 
$\theta=100^\circ$ in the $\theta$-ET family shows SC in place of CO below 
$T \approx 3.6$ K.~\cite{Kobayashi86} 

Thus far, the CO pattern in each compound has been investigated by X-ray 
diffraction and $^{13}$C-NMR measurements.  In the series of 
X$=MM^\prime$(SCN)$_4$ with $M^\prime=$Co and Zn, $M=$Tl and Rb, 
a stripe-type CO which has a twofold periodicity along the $c$ axis is 
observed below $T_{\rm CO}$.~\cite{Miyagawa00,Chiba01,Watanabe03,Watanabe04}  
The compounds with $M=$Tl and Rb is known to have an average dihedral angle 
of $\theta = 120^\circ$ to $110^\circ$.  Using the estimated ratio of the 
transfer integrals $t_c/t_p \sim 0.5-0.4$, the stripe-type CO was successfully 
reproduced by the mean-field theory.~\cite{Seo00}  But then, the X-ray diffuse 
scattering experiments~\cite{Nogami99,Watanabe99} demonstrated the presence 
of peculiar charge fluctuations in $\theta$-(ET)Cs$M^\prime$(SCN)$_4$ with 
$M=$Co and Zn; two diffuse peaks with the wave numbers 
${\bf q}_1=(q_a,q_b,q_c)=(0,k,1/2)$ and ${\bf q}_2=(2/3,k,1/3)$ were found.  
The wave number ${\bf q}_1$ is the same as that of the stripe-type CO observed 
in the compounds with $M=$Tl and Rb, but the other wave number ${\bf q}_2$ 
corresponds to a three-fold periodicity.  The compounds with $M=$Cs have 
relatively small dihedral angle $\theta = 104^\circ$ and are located in the 
vicinity of the quantum critical point $T_{\rm CO} \sim 0$ K.  Accordingly, 
the ratio of the transfer integrals is expected to be rather small, 
i.e., $t_c/t_p \ll 1$, which is different from the case of $M=$Tl and Rb.  
Thus, a very absorbing subject, i.e., the coexistence of charge fluctuations 
with different wave numbers, has been provided.  

The appearance of the spatially inhomogeneous CO has also been suggested 
by the dielectric response, ac resistivity,~\cite{Inagaki04} $^1$H-NMR, 
EPR, and static magnetic susceptibility~\cite{Nakamura00,Chiba03} measurements.  
Furthermore, a power-law behavior of the current-voltage characteristics 
over a wide range of currents~\cite{Takahide06} and a current-induced melting 
of insulating CO domains~\cite{Sawano05} have been observed.  In addition, 
the optical conductivity spectra show a transition from short-ranged to 
long-range CO around $T=T_{\rm CO}$, and stronger insulating features appear 
in $MM^\prime=$CsZn than in $MM^\prime=$RbZn.~\cite{Wang03,Suzuki05}  
A number of theoretical studies have then been carried out in order to 
clarify the nature of the CO and charge 
fluctuations.~\cite{Kaneko06,Watanabe06,Hotta06,Kuroki06,Seo06,Udagawa07} 

Motivated by such developments in the field, we consider in this paper 
the extended Hubbard model at quarter filling defined on the anisotropic 
triangular lattices.  We employ the density-matrix renormalization group 
(DMRG) and Lanczos exact-diagonalization methods to investigate the 
electronic states of the model.  First, we calculate the hole density and 
double occupancy to determine the ground-state phase diagram.  
Next, the dynamical density-density correlation functions are calculated 
to study the low-energy charge excitations.  
We also calculate the single-particle excitation spectrum to elaborate 
the anomalous metallic states seen in the low-energy charge excitations 
of the three-fold CO phase.  
We finally obtain the optical conductivity for each phase and clarify the 
behavior of the charge degrees of freedom of the model by focussing on the 
three-fold charge fluctuations.  

This paper is organized as follows. In Sec.~II, the 2D extended Hubbard model 
on the anisotropic triangular lattices is introduced.  We also define some 
physical quantities and explain the applied methods for the calculations.  
In Sec.~III, we present the ground-state phase diagram of the extended 
Hubbard model and discuss the physical properties related to the CO in each 
phase.  Section IV contains conclusions and discussions including comparison 
with the experimental results.

\section{Model and Method}

\subsection{Hamiltonian}

\begin{figure}[t]
\includegraphics[width= 7.0cm,clip]{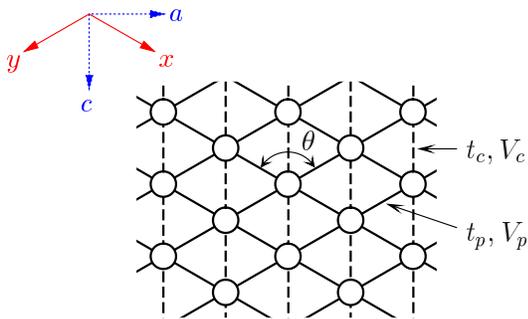}
\caption{(Color online) Schematic representation of the anisotropic 
triangular-lattice Hubbard model.  Each circle denotes an ET molecule.}
\label{lattice}
\end{figure}

The Hamiltonian of the extended Hubbard model defined on the anisotropic 
triangular lattice is given by
\begin{eqnarray}
\nonumber
H=&-&\sum_{\left\langle ij\right\rangle \sigma}t_{ij}
(c_{i\sigma}^\dag c_{j\sigma}+{\rm H.c.})\\
&+&U\sum_i n_{i\uparrow}n_{i\downarrow}
+\sum_{\left\langle ij\right\rangle}V_{ij}n_in_j
\label{hamiltonian}
\end{eqnarray}
where $c_{i\sigma}^\dag$ ($c_{i\sigma}$) is the creation (annihilation) 
operator of a hole with spin $\sigma$ at site $i$, 
$n_{i\sigma}=c_{i\sigma}^\dag c_{i\sigma}$ is the number operator, 
and $n_i=n_{i\uparrow}+n_{i\downarrow}$.  
The sum $\left\langle ij \right\rangle$ runs over nearest-neighbor pairs.  
$t_{ij}$ and $V_{ij}$ are hopping integral and intersite Coulomb repulsion 
between sites $i$ and $j$; here we retain two kinds of nearest-neighbor 
hopping integrals (repulsions) $t_p$ ($V_p$) and $t_c$ ($V_c$) as shown 
in Fig.~\ref{lattice}.  $U$ is the onsite Coulomb repulsion.  We restrict 
ourselves to the case at quarter filling; i.e., 
$\left\langle n_i \right\rangle = 0.5$ where 
$\left\langle \cdots \right\rangle$ denotes the 
ground-state expectation value.  
In this paper, we consider the case of $U/t_p=10$ and $t_c/t_p = 0$ as 
a typical set of parameter for X=CsCo(SCN)$_4$.~\cite{Mori98}  
Hereafter, we take $t_p = 1$ as the unit of energy.  

\subsection{Physical quantities}

We employ two kinds of numerical methods, i.e., the DMRG and 
Lanczos methods.  Either of these methods is chosen for the 
calculation of each physical quantity.  
The details are explained in the following.  

\subsubsection{Static quantities}

We are interested in the CO phenomena in the subject materials, 
so that it is useful, first of all, to make the ground-state 
phase diagram associated with the charge degrees of freedom.  
To this end, we calculate the hole density 
$\left\langle n_i\right\rangle$ and double occupancy 
$\left\langle d_i \right\rangle$ ($d_i=n_{i\uparrow}n_{i\downarrow}$) 
for all sites $i$ to investigate the charge distribution.  
A large size system is required for reproducing all possible CO 
patterns and for an accurate estimation of these quantities.  
We thus study a finite-size cluster of $L_a=8$ and $L_c=6$ using 
the DMRG method.  
The periodic boundary conditions (PBC) are applied for the $c$-direction, 
whereas the open-end boundary conditions (OBC) are applied for 
the $a$-direction.  This choice of the boundary conditions enables 
us to detect the CO state clearly (see Sec.III A).  We keep up to 
$m \simeq 1400$ density-matrix eigenstates in the DMRG procedure 
and thus the maximum truncation error, i.e., the discarded weight, 
is less than $1 \times10^{-4}$.  In this way, the maximum error in 
the ground-state energy is estimated to be $\Delta E_0/N \sim 10^{-2}$ 
for the most inaccurate case $V_p=V_c=0$.  Based on the results of 
the hole density and double occupancy, we determine the ground-state 
phase diagram on the parameter space $(V_p, V_c)$.  

\subsubsection{Dynamical density-density correlation}

Next, we consider the dynamical density-density correlation function 
defined as 
\begin{eqnarray}
N({\bf q},\omega)=\sum_\nu |\langle \psi^{N_h}_\nu |\hat{n}_{\bf q}| 
\psi^{N_h}_0 \rangle|^2 \delta(\omega-E_\nu+E_0) 
\label{Nqw}
\end{eqnarray}
in order to study the low-energy charge excitations.  
Here, $| \psi^{N_h}_\nu \rangle$ and $E^{N_h}_\nu$ are the $\nu$-th 
eingenstate and eigenenergy of the system with $N_h$ holes 
($\nu=0$ corresponds to the ground state).  
Since the exact definition of the momentum-dependent operators with 
the OBC is quite difficult, we choose the PBC for a quantitative 
estimation of the spectrum; we therefore use the PBC for both the 
$x$- and $y$-directions (or the $a$- and $c$-directions).  
With this PBC, the density operators $\hat{n}_{\bf q}$ can be precisely 
defined by 
\begin{eqnarray}
\hat{n}_{\bf q}=\frac{1}{\sqrt{N}}\sum_ie^{i{\bf q}\cdot{\bf r}_i}n_i 
\label{Nq}
\end{eqnarray}
where $N$ is the number of lattice sites $L_x\times L_y$ 
(or $L_a\times L_c$) and ${\bf r}_i$ is the position of site $i$. 
However, in this case, it is quite difficult 
to carry out sufficiently accurate calculations with the DMRG method, 
so that we here use the Lanczos method on small clusters with 
16 ($L_x=4$, $L_y=4$) and 18 ($L_a=6$, $L_c=3$) sites.  
Consequently, the system size is restricted but the result is 
numerically exact.  

\subsubsection{Single-particle excitation spectrum}

Of interest are also the evolution of the fundamental excitations 
in the systems with strongly frustrated correlations.  To see this, 
we calculate the single-particle excitation spectrum, which is 
obtained as
\begin{eqnarray}
A({\bf q},\omega) = A^-({\bf q},-\omega) + A^+({\bf q},\omega)
\label{Aqw}
\end{eqnarray}
with the photoemission (PES) spectrum
\begin{eqnarray}
A^-({\bf q},\omega) = \sum_{\nu \sigma}
|\langle \psi^{N_h-1}_\nu |\hat{c}_{{\bf q}\sigma}| 
\psi^{N_h}_0 \rangle|^2 \delta(\omega-E^{N_h-1}_\nu+E^{N_h}_0)
\label{PES}
\end{eqnarray}
and inverse photoemission (IPES) spectrum
\begin{eqnarray}
A^+({\bf q},\omega) = \sum_{\nu \sigma}
|\langle \psi^{N_h+1}_\nu |\hat{c}^\dagger_{{\bf q}\sigma}| 
\psi^{N_h}_0 \rangle|^2 \delta(\omega-E^{N_h+1}_\nu+E^{N_h}_0)
\label{IPES}
\end{eqnarray}
where $\hat{c}^\dagger_{{\bf q}\sigma}$ ($\hat{c}_{{\bf q}\sigma}$) 
is the Fourier transform of the creation (annihilation) operator 
$\hat{c}^\dagger_{i \sigma}$ ($\hat{c}_{i \sigma}$).  For the 
same reason as the calculation of the dynamical density-density 
correlation function, we employ the Lanczos method with the PBC 
where the operator is defined by
\begin{eqnarray}
\hat{c}_{{\bf q}\sigma}=\frac{1}{\sqrt{N}}\sum_i 
e^{i{\bf q}\cdot{\bf r}_i}\hat{c}_{i\sigma}. 
\label{cFT}
\end{eqnarray}
The calculation is carried out with a 18-site ($L_a=6$, $L_c=3$) 
cluster.

\subsubsection{Optical conductivity}

Finally, for more elaborate study on the behavior of the charge 
degrees of freedom and direct comparison with experiments, we calculate 
the optical conductivity defined by
\begin{eqnarray}
\sigma_\alpha(\omega)=\frac{\pi}{N}\sum_\nu \omega 
|\langle\psi_\nu |\hat{D}_\alpha| 
\psi_0 \rangle|^2 \delta(\omega-E_\nu+E_0),
\label{sigma}
\end{eqnarray}
with a component of the dipole operator
\begin{eqnarray}
\hat{D}_\alpha=-e\sum_i{\bf r}_i\cdot{\bf v}_\alpha(n_i-\frac{1}{2}),
\label{dipole}
\end{eqnarray}
where ${\bf v}_\alpha$ is a unit vector parallel to the $\alpha$-direction. 
It is known that the OBC is more feasible for calculation of the optical 
conductivity spectra because finite-size effects have much smaller influence 
on the results in comparison with a case using the PBC.~\cite{Moskvin03}  
Hence, we apply the OBC for both the $a$- and $c$-directions and study 
a finite-size cluster $L_a=8$, $L_c=6$ with the dynamical DMRG (DDMRG) method.  
The DDMRG method is an extension of the standard DMRG method for the 
calculation of dynamical properties.  In the DDMRG calculation, a required 
CPU time increases rapidly with the number of the density-matrix eigenstates 
$m$, so that we try to keep it as few as possible.  Because the DMRG approach 
is based on a variational principle, we have to prepare a `good trial function' 
of the ground state with the density-matrix eigenstates as much as possible.  
We therefore keep $m \sim 600$ to obtain the true ground state in the first 
five DDMRG sweeps and keep $m \sim 350$ to calculate the spectrum of 
the system.  As a result, the maximum truncation error, i.e., the discarded 
weight, is about $1\times10^{-3}$, while the maximum error 
in the ground-state is about $\Delta E_0/N\sim 2\times 10^{-2}$.  

\section{Results}

\subsection{Ground-state phase diagram}

\begin{figure}[t]
\includegraphics[width= 8.0cm,clip]{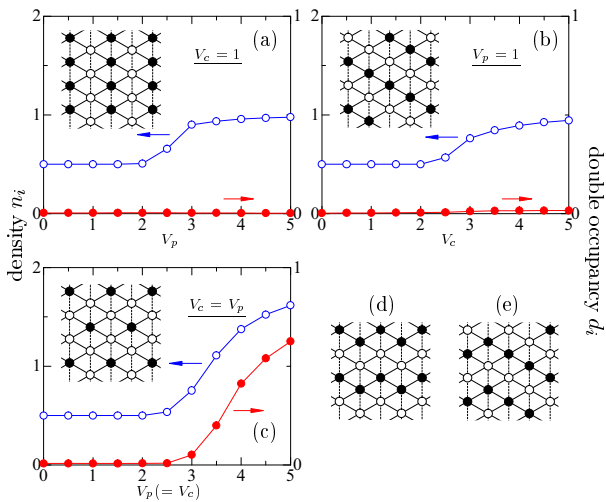}
\caption{(Color online) Hole density $\left\langle n_i\right\rangle$ and 
double occupancy $\left\langle d_i\right\rangle$ at the hole-rich sites for 
(a) $V_p=1$, (b) $V_c=1$, and (c) $V_p=V_c$.  The inset shows the schematic 
representation of (a) the diagonal stripe, (b) vertical stripe, and 
(c) three-fold CO patterns.  The solid circle denotes a site where the hole 
predominantly resides.  The other configurations are also shown: 
(d) horizontal and (e) an example of randomly aligned CO patterns.}
\label{n_d}
\end{figure}

So far, it has been suggested that the system defined by 
Eq.(\ref{hamiltonian}) has three types of CO phases, depending on the 
values of the intersite Coulomb repulsions.  The CO patterns 
for each phase are shown in the inset of Fig.~\ref{n_d}: 
(a) vertical stripe for $V_c \ll V_p$, 
(b) diagonal stripe for $V_c \gg V_p$, and 
(c) three-fold alignment for $V_c \sim V_p \gg t_p$.  
Other likely patterns are also shown: 
(d) horizontal and (e) randomly aligned stripes, 
which are not realized in the ground state.  A random alignment CO 
pattern is formed with any combination of alternately-occupied chains 
parallel to the $c$-axis.  Hence, the diagonal CO pattern is a 
special case of the random alignment ones. 

First of all, we note that the CO is observed as a state with a broken 
translational symmetry in our calculations: actually, there are two or 
more degenerate ground states and one of the degenerate states is picked 
out as the ground state by an initial condition of the DMRG calculation 
as we here apply the OBC along the $x$-direction.  Thus, a CO state 
includes more than two kinds of sites with different hole densities, 
i.e., hole-rich and hole-poor sites.  
In Fig.~\ref{n_d}, we show DMRG results of the hole density 
$\left\langle n_{i \in {\rm rich}} \right\rangle$ and double occupancy 
$\left\langle d_{i \in {\rm rich}} \right\rangle$ at the hole-rich sites 
$i \in {\rm rich}$ for (a) $V_c=1$, (b) $V_p=1$, and (c) $V_p=V_c$.  
When both $V_p$ and $V_c$ are small, the analysis 
shows that the hole density is uniform over the system, i.e., 
$\left\langle n_i\right\rangle=0.5$ for all sites $i$, and the system 
is in the metallic state.  
As $V_p$ and/or $V_c$ increase, the charge fluctuations toward the CO 
states are enhanced and we find that 
$\left\langle n_{i\in{\rm rich}}\right\rangle$ starts to increase 
at the CO phase boundary.  In the CO phase, the charge fluctuations 
are weakened again as the CO state is stabilized, where 
$\left\langle n_{i\in{\rm rich}}\right\rangle$ approaches a value in 
the atomic limit $t_p\to 0$.  

Let us now evaluate the phase boundaries between the metallic and CO 
phases.  When we increase $V_p$ from zero with fixing $V_c=1$ 
[see Fig.~\ref{n_d}(a)], $\left\langle n_{i\in{\rm rich}}\right\rangle$ 
begins to increase around $V_p=2.5$; accordingly, a critical strength 
of the vertical CO state is obtained as 
$(V_p^{\rm c},V_c^{\rm c})\sim(2.5,1)$.  
At $V_p>V_p^{\rm c}$, $\left\langle n_{i\in{\rm rich}}\right\rangle$ 
increases rapidly to unity but the double occupancy remains almost zero.  
This means that (nearly) half-filled chains are alternated with 
(nearly) empty chains and the charge fluctuations are totally 
suppressed in the entire region of vertical CO phase.  
In analogy with the case of the vertical CO state, we can estimate 
a critical strength of the diagonal CO state as 
$(V_p^{\rm c},V_c^{\rm c})\sim(1,2.75)$ [see Fig.~\ref{n_d}(b)].  
We should however note that $\left\langle n_{i \in{\rm Rich}}\right\rangle$ 
seems to approach unity more slowly than that in the vertical CO phase.  
This implies that the charge fluctuations in the  diagonal CO state 
are rather stronger than those in the vertical CO state.  The reason of 
this strong charge fluctuation in the diagonal CO phase is the 
existence of nearly energetically degenerate states, i.e., the horizontal 
and an infinite number of the randomly aligned CO patterns, which are 
shown in Fig.~\ref{n_d}(d) and (e), respectively.  
This situation has also been confirmed by the mean-field 
approximation,~\cite{Kaneko06} variational Monte Carlo (VMC) 
method,~\cite{Watanabe06} and strong-coupling study of the spinless 
model.~\cite{Hotta06}  The energies of those CO patterns including the 
diagonal one are exactly equal to each other in the atomic limit $t_p \to 0$.

\begin{figure}[t]
\includegraphics[width= 8.5cm,clip]{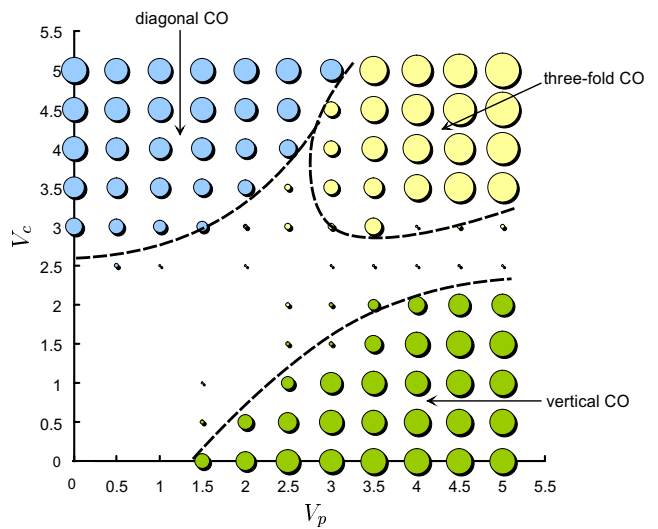}
\caption{(Color online) Ground-state phase diagram of the extended 
Hubbard model defined on the anisotropic triangular lattice.  
We assume the values $t_{\rm c} = 0$ and $U = 10$.  
The result is obtained from the DMRG calculations for the charge 
distributions with a broken translational symmetry.  
The diameter of the circles is taken to be proportional 
to the hole density at the hole-rich sites.}
\label{PD}
\end{figure}

As for the three-fold CO phase, the situation is somewhat different 
from the other CO phases.  If a relation $V_c=V_p$ is kept, the 
three-fold CO instability begins to appear around a critical point 
$(V_p^{\rm c},V_c^{\rm c}) \sim (3,3)$ [see Fig.~\ref{n_d}(c)].  
Above the critical point, we find the rapid increase of not only 
the hole density but also the double occupancy at the hole-rich sites.  
This implies that the competition between the effects of $U$ and $V$'s 
is essential for the presence of the three-fold CO phase.  We also 
note that the three-fold CO state is metallic in the sense that the 
Drude weight is nonzero.~\cite{Shingai}  The presence of metallic CO 
phase has previously been suggested for the extended square-lattice 
Hubbard models; simply, the pocket-like small Fermi surface appears 
by doping the long-range CO phases with holes or electrons without 
destroying the long-range CO.~\cite{Ohta94} 
Note that the ground state has both the onsite hole-pairing and CO, 
which may be essentially the same as the coexistence of the 
{\it s}-wave superconducting and CO states in the two-dimensional 
negative-$U$ Hubbard model.~\cite{Scalettar89}

\begin{figure}[t]
\includegraphics[width= 7.0cm,clip]{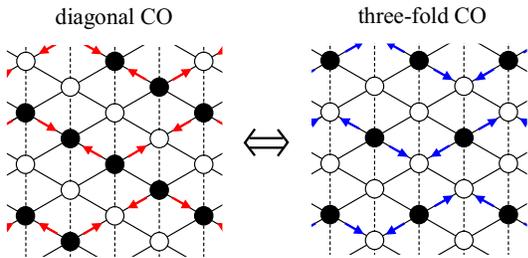}
\caption{(Color online) Transition between diagonal and three-fold 
CO states.  The solid circle denotes a site where hole predominantly 
resides.  The arrows denote the flow of charges toward the other 
CO state.}
\label{flow}
\end{figure}

In Fig.~\ref{PD}, we determine the ground-state phase diagram from 
the results of the hole density, where the phase boundaries are drawn 
with dashed curves.  We can make it certain that the three substantial 
CO phases are distinguished from a metallic phase with uniform charge 
distribution.  This phase diagram is basically consistent with that 
from the VMC calculation.~\cite{Watanabe06}  The main difference is 
that a metallic region is sandwiched between the vertical and 
three-fold CO phases in our results.  On the other hand, the diagonal 
and three-fold CO phases are contiguous, in agreement with the VMC 
result.  The appearance of the metallic region in-between may be 
explained as follows.  The diagonal and three-fold CO phases may 
be contiguous because the transition can be derived by the flow of 
charges only via the nearest-neighbor hopping integrals $t_p$ 
(see Fig.~\ref{flow}), whereas a drastic charge redistribution is 
required in the transition from the vertical CO pattern to the 
three-fold CO pattern and therefore the ground state needs to go 
through the metallic one.  
We may also suggest that this `charge-flowing' transition between 
the diagonal and three-fold CO phases occurs in the presence of the 
three-fold (diagonal) charge fluctuations in the diagonal (three-fold) 
CO phase as is evident in the calculated dynamical density-density 
correlation functions (see the following subsection).  
We also note that the three-fold CO phase is shrunk with increasing 
$U$.

\subsection{Dynamical density-density correlation}

\begin{figure}[t]
\includegraphics[width= 8.5cm,clip]{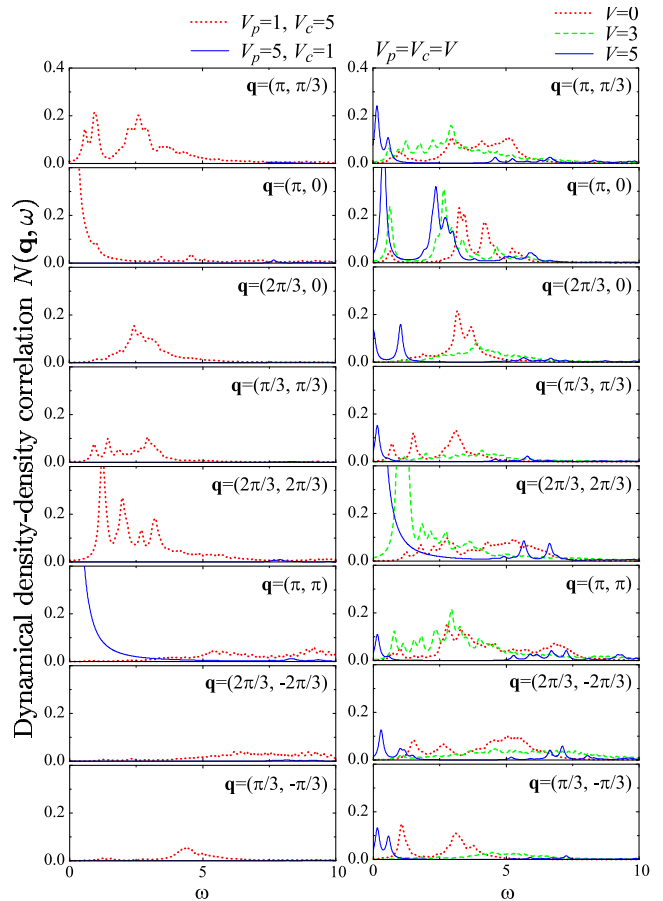}
\caption{(Color online) Charge excitation spectra calculated for 
the 18-site cluster with the Lanczos exact-diagonalization method.  
Only the results of $(\pi,0)$ are obtained for the 16-site cluster.  
Spectra at $(V_p,V_c)=(1,5)$, $(5,1)$ [left panel], $(0,0)$, 
$(3,3)$, and $(5,5)$ [right panel] are shown.}
\label{nqw}
\end{figure}

The dynamical density-density correlation functions calculated 
by the Lanczos method for several sets of $V_p$ and $V_c$ values 
are shown in Fig.~\ref{nqw}.  Each parameter set corresponds to a 
different phase as follows: $(V_p,V_c)=(1,5)$ to the diagonal CO 
phase, $(5,1)$ to the vertical CO phase, $(5,5)$ to the three-fold 
CO phase, and $(0,0)$ to the uniform metallic phase.  
In addition, we plot the results for $(V_p,V_c)=(3,3)$, where 
the system is in the non-CO metallic phase but with strong three-fold 
CO fluctuations.  
When $(V_p,V_c)=(0,0)$, the model (\ref{hamiltonian}) is equivalent 
to the square-lattice Hubbard model at quarter filling and no 
particular modes of charge fluctuations are developed; we therefore 
find the broad spectral features for all the momenta, which come 
basically from the particle-hole transition in the corresponding 
noninteracting system.  With increasing the intersite interactions, 
however, the broad features change to the low-energy sharp peaks, 
indicating that the excitations turn to be collective-mode like.  

Generally, the enhancement of low-lying peak at a particular momentum 
must be expected for the CO instability.  When the system is unstable 
to the vertical [diagonal] CO phase, the particular momentum is located 
at ${\bf q}=(\pi,\pi)$ [${\bf q}=(\pi,0)$].  Looking at the spectra 
for $(V_p,V_c)=(5,1)$, a sharp peak appears around $\omega=0$ at 
${\bf q}=(\pi,\pi)$ and there are nearly no peaks at the other momenta.  
This means that the system is unstable to the vertical CO state with 
(almost) complete charge disproportionation.  This result is consistent 
with the results of the hole density given in Sec.~III A.  
On the other hand, in the spectra at $(V_p,V_c)=(1,5)$, we see not 
only a low-energy peak at $(\pi,0)$ but also relatively high-energy 
peaks at the other momenta; most of the high-energy peaks concentrate 
at momenta with $q_xq_y>0$ and few peaks appear at momenta with $q_xq_y<0$.  
By the axes rotation, we find the charge fluctuations mostly occur 
along the $c$-axis, which are associated with the nearly degenerate 
states, i.e., horizontal and randomly aligned CO phases.  It is moreover 
striking that the intensity around ${\bf q}=(2\pi/3,2\pi/3)$ is 
significantly large, which means that the three-fold charge fluctuations 
are rather strong although the system is in the diagonal CO phase.  

Let us now discuss the spectral features for the three-fold CO metallic 
state, of which the particular momentum is located at 
${\bf q}=(2\pi/3,2\pi/3)$.  
The charge excitation spectra in the CO metallic state are currently 
not well understood, so that a good opportunity for studying them 
may be offered here.  We thus investigate a process of gradual change 
in the spectra with increasing the intersite Coulomb interactions $V$'s.  
Since the three-fold CO phase lies around $V_p\sim V_c>3$, we increase 
$V_p$ ($V_c$) from $0$ to $5$ with keeping a condition $V_p=V_c$.  
The spectra for $(V_p,V_c)=(0,0)$ have the broad spectral features for 
all the momenta as mentioned above and these features basically remain 
unchanged as far as $V_p$ and $V_c$ are less than $\sim 2$.  
Only around $(V_p,V_c)=(3,3)$, we can recognize the changes clearly in 
the calculated spectra: the low-energy spectrum at ${\bf q}=(2\pi/3,2\pi/3)$ 
is particularly enhanced as anticipated, but those at ${\bf q}=(\pi,0)$ 
and ${\bf q}=(\pi,\pi)$ are also enlarged, while the spectral weights 
for the other momenta are strongly suppressed.  We understand that these 
changes suggest that the three types of CO instability are simultaneously 
developed and that, considering that the energies of the lowest-lying 
excitations for the three momenta are nearly equal, the charge fluctuations 
for the three CO patterns are strongly competing.

After increasing the intersite Coulomb interactions to $(V_p,V_c)=(5,5)$, 
even greater changes are seen in the spectral features.  
At ${\bf q}=(2\pi/3,2\pi/3)$, we find the enhancement of a sharp peak 
around $\omega=0$, which indicates the strong charge fluctuations of 
the three-fold periodicity and is consistent with the results of the hole 
density.  It is particularly worth noting that the low-energy spectral 
intensities at $(\pi,0)$ is continuously enhanced [and those at 
$(\pi,\pi)$ are suppressed].  

Let us discuss some experimental situations here.  
We may point out first that the wave vectors ${\bf q}_1$ and ${\bf q}_2$ 
of the two diffuse peaks observed experimentally~\cite{Nogami99,Watanabe99} 
in the CO phase of $\theta$-(ET)$MM^\prime$(SCN)$_4$ [$M=$Cs, $M=$Co or Zn] 
should be equivalent to our momenta $(\pi,0)$ and $(2\pi/3,2\pi/3)$, 
respectively.  
We may therefore suggest that the appearance of the two enhanced peaks 
in our calculations correspond to the coexistence of the two different 
charge fluctuations.  Note that these enhanced peaks are already obtained 
for more realistic parameter values $(V_p,V_c)=(3,3)$ in our calculations.  
It is also striking that the low-energy peaks appear in the entire 
Brillouin zone and the charge fluctuations are induced in all the momentum 
transfers.  This situation is quite different from those for the diagonal 
and vertical CO stripes where the low-energy peak appears only at the 
particular momenta.  

\subsection{Single-particle excitation spectrum}

\begin{figure}[t]
\includegraphics[width= 8.0cm,clip]{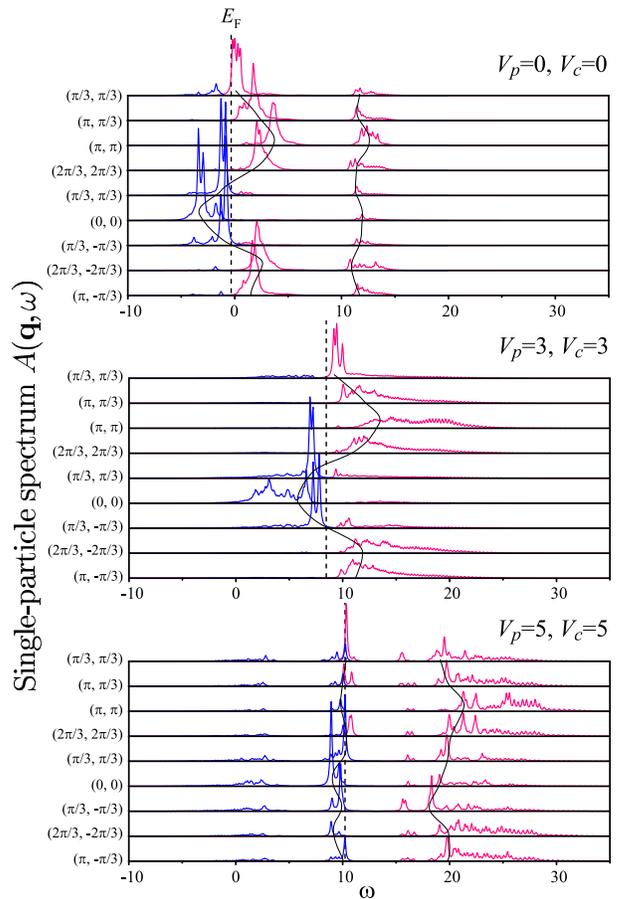}
\caption{(Color online) Single-particle excitation spectra calculated 
for the 18-site cluster with the Lanczos exact-diagonalization method.  
The dashed lines denote the Fermi levels and the thin lines mark 
the dispersion of bands.}
\label{Aqwspec}
\end{figure}

In the diagonal and vertical CO phases, we just find the insulating 
band structures reflecting the CO states in our calculations; namely, 
the PES and IPES peaks are separated by the charge gap and the Fermi 
level lies between the two spectra.  We thus focus on the spectral 
features in the three-fold CO phases here.  In Fig.~\ref{Aqwspec}, 
we show the single-particle excitation spectra calculated with the 
Lanczos method for $(V_p,V_c)=(0,0)$, $(3,3)$, and $(5,5)$.  
At $(V_p,V_c)=(0,0)$, we can clearly see the spectral weight separated 
into two bands due to the onsite Coulomb interaction, i.e., lower 
Hubbard band (LHB) and upper Hubbard band (UHB) with the distance 
$\sim 10$ ($=U$).  The LHB, which corresponds to the conduction band, 
can be approximately expressed by the dispersion of noninteracting 
holes,
\begin{eqnarray}
\varepsilon_{\bf q}=-2t_p \cos q_x -2t_p \cos q_y.
\label{bareband}
\end{eqnarray}
On the other hand, the UHB has relatively small spectral weight and 
narrow band width, which are general features in the strong-coupling 
regime.  At $(V_p,V_c)=(3,3)$, the shape of conduction band remains 
almost unchanged from that at $(V_p,V_c)=(0,0)$.  However, we can see 
that an overlap between the LHB and UHB occurs due to the competition 
between the onsite and nearest-neighbor Coulomb interactions.  
As a result, the band structure is seen to be reduced to a single 
dispersion line.  The interaction between holes is effectively weakened 
(or vanished) by the frustration of the Coulomb interactions, so that 
the dispersion relation looks as if it were for the single-orbital 
model or the noninteracting case.  

Surprisingly, the spectral features are drastically changed at 
$(V_p,V_c)=(5,5)$.  The spectral weight appears to be separated to 
form several bands again.  It is particularly worth noting that 
the width of conduction band is much narrower than that at 
$(V_p,V_c)=(3,3)$.  In the three-fold CO phase, the holes can conduct 
only through the hole-rich sites but the hole-rich sites are not 
connected by direct hopping integrals; in other words, the carriers 
form a small Fermi surface.~\cite{Ohta94}  Thus, the conduction band 
is centered at $\omega \sim 10 (=U)$ and the band width should be very 
narrow.  It may be related to the appearance of low-energy peaks in 
the entire Brillouin zone observed in the dynamical density-density 
correlation function as discussed above.  Let us then consider the 
peaks in the higher-energy range, $\omega \gtrsim 15$.  These peaks 
are essentially formed in connection with the empty (hole-poor) 
sites.  An empty site is surrounded by three hole-rich sites and a 
hole-rich site is occupied by one or two holes.  When a hole is added 
on an empty site, the excitation energy is $\omega \sim 2V_p+V_c=15$ 
if the surrounding three hole-rich sites are singly occupied and 
it is $\omega \sim 4V_p+2V_c=30$ if the surrounding three hole-rich 
sites are doubly occupied.  This is the reason why there are continuum 
weights from $\omega \sim 15$ to $30$ in the spectra for all the momenta.  
In addition, we can find the dispersion with bandwidth $\sim 4$ around 
$\omega=20$.  This dispersion is formed from a empty zigzag chain along 
the $a$-axis, which is nearly separated from each other by the 
hole-rich sites.  

\subsection{Optical conductivity}

\begin{figure}[t]
\includegraphics[width= 8.5cm,clip]{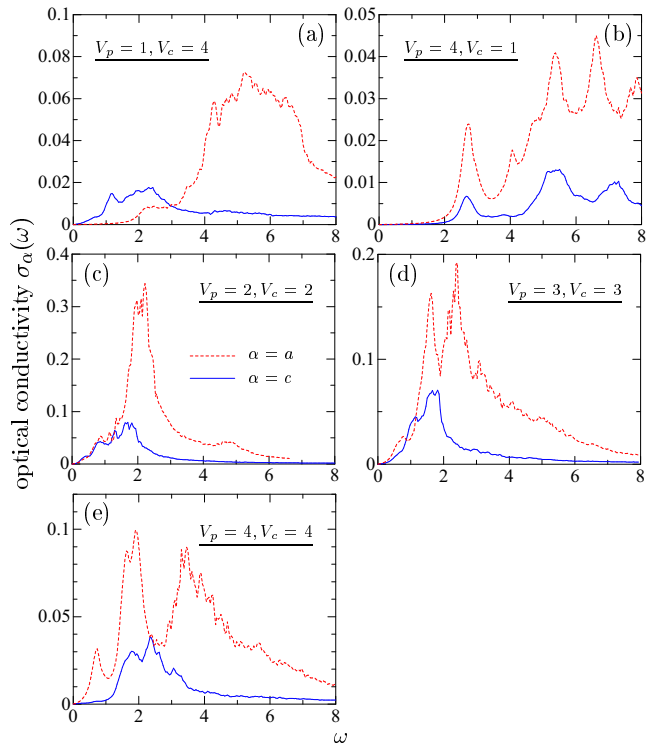}
\caption{(Color online) Optical conductivity spectra calculated 
with the DDMRG method.  $\alpha (=a,c)$ is the direction of 
the electric field defined in Fig.~\ref{lattice}.}
\label{opt}
\end{figure}

The optical conductivity spectra calculated by the DMRG method for 
several sets of $V_p$ and $V_c$ are shown in Fig.~\ref{opt}.  
For all cases, the spectral intensity for the $a$-direction seems 
to be a few times larger than that for the $c$-direction, which 
just reflects the difference between the interatomic distances 
$d_a:d_c=\sqrt{3}:1$ in the $a$ and $c$ directions.  This leads 
to the intensity ratio $I_a:I_c=3:1$. 

The most easily comprehensible result should be the ones in the 
vertical CO phase, i.e., $(V_p,V_c)=(4,1)$ [see Fig.~\ref{opt}(b)].  
The shapes of spectra for the $a$- and $c$-directions are essentially 
the same, although the intensities are different as mentioned above.  
For both of the spectra, the lowest excitation appears around 
$\omega\sim 2.5$, which corresponds to the optical (insulating) gap, 
and the other excitations lie in much higher frequencies, 
$\omega\gtrsim 4$.  Also, the total weight of the spectrum is smaller 
than that in the other CO phases.  These results are consistent 
with the fact that the vertical CO state is rather solid and the 
charge fluctuations are quite small.  The optical gap in the vertical 
CO phase may be estimated as $\sim 3V_p-2V_c-W$ with the bare bandwidth 
$W=8$.

Let us then turn to the diagonal CO phase, i.e., $(V_p,V_c)=(1,4)$ 
[see Fig.~\ref{opt}(a)].  We find that the result for the $a$-direction 
is apparently different from that for the $c$-direction.  
The spectrum for the $a$-direction consists of the first-excitation peak 
at $\omega \sim 2.2$ and high-energy broad features, which is similar 
to that in the vertical CO phase.  This result reflects the solid CO 
and weak charge fluctuations along the $a$-direction, as shown in the 
dynamical density-density correlation functions.  On the other hand, 
most of the spectral features for the $c$-direction appear in the 
lower-energy range.  The optical gap is of the order of $\sim 0.5$, 
which is much smaller than that of the vertical CO phase.  This is so 
because the diagonal CO state can be rearranged to the randomly aligned 
(or horizontal) CO state with small excitation energy.  Thus, the charge 
fluctuations along the $c$-axis are much stronger than those along 
the $a$-axis.  Since the energy difference between the diagonal CO phase 
and random alignment CO state is hardly changed even if $V_c$ increases, 
the low-energy structure is expected to be seen for any $V_c$ value.

Lastly, we discuss the spectral aspects in the metallic regime around 
the three-fold CO instability [see Fig.~\ref{opt}(c)-(e)].  With the OBC, 
the optical conductivity at $\omega=0$ is ruled out and the lowest-energy 
scale is limited to an order of $4W/L_aL_b \sim 0.67$.~\cite{Fye91}  
Therefore, the Drude spectral weight appears around the minimum excitation 
energy.  At $(V_p,V_c)=(2,2)$, the spectra for the $a$- and $c$-directions 
are quite similar.  For each spectrum, small peak structure around 
$\omega\sim 0.7-0.8$ should correspond to the Drude peaks.  The spectral 
weight exists continuously from the `Drude frequency' to higher 
excitations and reaches the maximum around $\omega=2$, and very little 
spectral weight exists at higher-energy range, $\omega \gtrsim 3$.  
At $(V_p,V_c)=(3,3)$, we can find some tendencies to the three-fold CO 
state; in the spectrum along the $a$-axis the intensity around $\omega=2$ 
is about half of that at $(V_p,V_c)=(2,2)$ and the high-energy broad 
features are relatively large.  A large peak around $\omega=2$ at 
$(V_p,V_c)=(2,2)$ splits into two peaks that indicate the appearance of 
individual excitations associated with the COs.  The spectrum along the 
$a$-axis remains almost unchanged, except the Drude-like peak which becomes 
much smaller.  At $(V_p,V_c)=(4,4)$, we can clearly see the characteristic 
features of the three-fold CO state: in the spectrum along the $a$-axis 
the weight is separated into the low-energy Drude-like peak at 
$\omega\sim 0.7-0.8$ and high-energy broad features.  We can understand 
that the former comes from the small Fermi surface in the three-fold CO 
phase and the latter comes from the excitations due to the three-fold charge 
fluctuations.  On the other hand, the weight of the Drude-like peak is 
nearly zero in the spectrum along the $c$-axis.  This means that the CO 
is more solid along the $c$-axis than along the $a$-axis.  We can explain 
this as follows: 3/2-filled chains are alternated with zigzag empty chains 
along the $c$-axis [see the inset of Fig.~\ref{n_d} (c)], whereas each 
chain along the $c$-axis is quarter filled, which means that the holes 
distribute uniformly along the $a$-axis.  As mentioned above, the optical 
conductivity along the $a$- ($c$-) axis is concentrated in the lower-energy 
range when the three-fold (diagonal) charge fluctuations are enhanced.  
We suggest that these features in the optical conductivity spectra 
should give a good criterion to examine the CO patterns experimentally.  

\section{Conclusions}

The extended Hubbard model at quarter filling defined on the anisotropic 
triangular lattice have been studied by the Lanczos exact-diagonalization 
and DMRG methods.  We determine the ground-state phase diagram based on 
the results of the hole density and double occupancy.  In the phase 
diagram, there exist three substantial CO phases (diagonal, vertical, and 
three-fold) and a metallic phase with uniform charge distribution.  
We find that the charge fluctuations (or instability to CO) with the 
three-fold periodicity appear for realistic parameters 
$V_p\sim V_c\sim 0.3U$.  

We suggest that the transition between the diagonal and three-fold CO 
phases is derived by the flow of charges only via the nearest-neighbor 
hopping integrals, which may lead to the coexistence of the diagonal and 
three-fold charge fluctuations near the phase boundary.  The coexistence 
features are also found in the density-density correlation functions; 
in the diagonal (three-fold) phase, not only the particular peak but also 
the low-energy peaks corresponding to the three-fold (diagonal) charge 
fluctuations are enhanced.  In the vertical CO phase, only the low-energy 
peak at the particular momentum is visible and there is no charge 
fluctuations associated with the other CO states.  Moreover, we find that 
there appear the low-energy collective-modes-like excitations in the entire 
Brillouin zone when the three-fold charge fluctuations are very strong.  
The wave numbers of the diagonal and three-fold charge fluctuations are 
equivalent to those of the two X-ray diffuse peaks ${\bf q}_1=(0,k,1/2)$ 
and ${\bf q}_2=(2/3,k,1/3)$ measured in $\theta$-(ET)Cs$M^\prime$(SCN)$_4$ 
with $M=$Co or Zn.  If we could make the dihedral angle larger, the value 
of $V_p/V_c$ decreases and only the diagonal CO fluctuation remains.  
This is consistent with the fact that only a diffuse peak with wave number 
${\bf q}_1$ is observed in the compounds with $M=$Tl and Rb which have a 
larger dihedral angle than those with $M=$Co or Zn do.  

We also study the single-particle excitation spectrum to see the evolution 
of the fundamental low-lying excitations in the presence of strongly 
frustrated correlations.  When the onsite and intersite Coulomb interactions 
compete ($V_p \sim V_c \sim U/3$), the interaction between holes is 
effectively diminished and the dispersion relation looks as if it is of 
the noninteracting case.  When the two intersite Coulomb interactions 
compete ($V_p \sim V_c > U/3$), the carriers form very narrow conduction 
band with a small Fermi surface.  

Furthermore, we examine the optical conductivity which reflects the 
characteristic features for each CO phase.  In the vertical CO phase, 
the spectra for both the directions explicitly represent the insulating 
features and the optical gaps are large, estimated as $\sim 3V_p-2V_c-8$.  
In the diagonal CO phase, the spectrum for the $a$-direction is 
essentially the same as that in the vertical CO phase but that for the 
$c$-direction is located in the lower-energy range.  This is so because 
the diagonal CO state can be rearranged to the randomly aligned (or 
horizontal) CO state with small excitation energy.  In the three-fold CO 
phase, the spectra indicate the presence of the separated low-energy 
Drude-like peak and high-energy broad features, leading to the anomalous 
metallic states in the system.  The optical conductivity along the $a$- 
($c$-) axis is concentrated in the lower-energy range when the three-fold 
(diagonal) charge fluctuations.  These spectral features may give a good 
criterion to examine the CO patterns experimentally.  
Finally, we make a comment on the difference in the experimental spectra 
between $MM^\prime=$CsZn and $MM^\prime=$RbZn.  Because we find that an 
increase in $V_p/V_c$ makes the diagonal CO state unstable, it is the 
smallness of $t_c$ that is essential for the strong insulating features 
in the spectra of $MM^\prime=$CsZn.  

We hope that our studies will help us understand the mechanism of charge 
fluctuations in the extended Hubbard model defined on the anisotropic 
triangular lattices and hence will offer useful suggestions to some aspects 
of the three-fold charge fluctuations observed in $\theta$-(ET)$_2$X.  

\acknowledgments

This work was supported in part by Grants-in-Aid for Scientific Research
(Nos. 18540338, 18028008, 18043006, and 19014004) from the Ministry of 
Education, Science, Sports, and Culture of Japan.  A part of computations 
was carried out at the Research Center for Computational Science, Okazaki 
Research Facilities, and the Institute for Solid State Physics, 
University of Tokyo.


\begin{thebibliography}{99}
\bibitem[$^\ast$]{}Present address: Leibniz-Institut f\"ur Festk\"orper- und Werkstoffforschung 
Dresden, P.O. Box 270116, D-01171 Dresden, Germany
\bibitem{Tokura00} Y. Tokura and N. Nagaosa, Science {\bf 288}, 462 (2000).
\bibitem{Verwey47} E. J. W. Verwey and P. W. Haayman, and F. C. Romeijn, 
J. Chem. Phys. {\bf 15}, 181 (1947).
\bibitem{Mori98-1} S.~Mori, C.~H.~Chen, and S.-W.~Cheong, Nature {\bf 392}, 473 (1998).
\bibitem{Tranquada95} J.~M.~Tranquada, B.~J.~Sternlieb, J.~D.~Axe, Y.~Nakamura, and 
S.~Uchida, Nature {\bf 375}, 561 (1995).
\bibitem{Lang03}  M.~Lang and J.~M\"uller, {\it The Physics of Superconductors - Vol.II}, 
edited by K.-H. Bennemann and J. B. Ketterson (Springer, Berlin, 2003), p.453.
\bibitem{Mori98} H.~Mori, S.~Tanaka, and T.~Mori, \prb {\bf 57}, 12023 (1998); 
\bibitem{Mori00} H.~Mori, T.~Okano, S.~Tanaka, M.~Tamura, Y.~Nishio, K.~Kajita, 
and T.~Mori, J. Phys. Soc. Jpn. {\bf 69}, 1751 (2000).
\bibitem{Kobayashi86} H.~Kobayashi, R.~Kato, A.~Kobayashi, Y.~Nishio, K.~Kajita, and W.~Sasaki, 
Chem. Lett. {\bf 1986}, 789 (1986); {\bf 1986} 833 (1986).
\bibitem{Miyagawa00} K.~Miyagawa, A.~Kawamoto, and K.~Kanoda, \prb {\bf 62}, R7679 (2000).
\bibitem{Chiba01} R.~Chiba, H.~Yamamoto, K.~Hiraki, T.~Takahashi and T.~Nakamura, 
J. Phys. Chem. Solids {\bf 62}, 389 (2001). 
\bibitem{Watanabe03} M.~Watanabe, Y.~Noda, Y.~Nogami, and H.~Mori, 
Synth. Met. {\bf 135-136}, 665 (2003). 
\bibitem{Watanabe04} M.~Watanabe, Y.~Noda, Y.~Nogami, and H.~Mori, 
J. Phys. Soc. Jpn. {\bf 73}, 116 (2004);{\bf 74}, 2011 (2005).
\bibitem{Seo00} H.~Seo, J. Phys. Soc. Jpn. {\bf 69}, 805 (2000).
\bibitem{Nogami99} Y.~Nogami, J.-P.~Pouget, M.~Watanabe, K.~Oshima, H.~Mori, S.~Tanaka 
and T.~Mori, Synth. Met. {\bf 103} 1911 (1999).
\bibitem{Watanabe99} M.~Watanabe, Y.~Nogami, K.~Oshima, H.~Mori, and S.~Tanaka, 
J. Phys. Soc. Jpn. {\bf 68}, 2654 (1999).

\bibitem{Inagaki04} K.~Inagaki, I.~Terasaki, H.~Mori, and T.~Mori, 
J. Phys. Soc. Jpn. {\bf 73}, 3364 (2004).
\bibitem{Nakamura00} T.~Nakamura, W.~Minagawa, R.~Kinami, and T.~Takahashi, 
J. Phys. Soc. Jpn. {\bf 69}, 504 (2000).
\bibitem{Chiba03} R.~Chiba, K.~Hiraki, T.~Takahashi, H.~M.~Yamamoto, and T.~Nakamura, 
Synth. Met. {\bf 133-134}, 305 (2003).
\bibitem{Sawano05} F.~Sawano, I.~Terasaki, H.~Mori, T.~Mori, M.~Watanabe, N.~Ikeda, 
Y.~Nogami, and Y.~Noda, Nature {\bf 437}, 522 (2005).
\bibitem{Takahide06} Y.~Takahide, T.~Konoike, K.~Enomoto, M.~Nishimura, T.~Terashima, 
S.~Uji, and H.~M.~Yamamoto, \prl {\bf 96}, 136602 (2006).
\bibitem{Wang03} N.~L.~Wang, T.~Feng, Z.~J.~Chen, and H.~Mori, 
Synth. Met. {\bf 135-136}, 701 (2003).
\bibitem{Suzuki05} K.~Suzuki, K.~Yamamoto, K.~Yakushi, and A.~Kawamoto, 
J. Phys. Soc. Jpn. {\bf 74}, 2631 (2005).

\bibitem{Kaneko06} M.~Kaneko and M.~Ogata, J. Phys. Soc. Jpn. {\bf 75}, 014710 (2006).
\bibitem{Watanabe06}  H.~Watanabe and M.~Ogata, J. Phys. Soc. Jpn. {\bf 75}, 063702 (2006).
\bibitem{Hotta06} C.~Hotta and N.~Furukawa, \prb {\bf 74} 193107 (2006); 
C.~Hotta, N.~Furukawa, A.~Nakagawa, and K.~Kubo, J. Phys. Soc. Jpn. {\bf 75}, 123704 (2006).
\bibitem{Kuroki06} K.~Kuroki, J. Phys. Soc. Jpn. {\bf 75}, 114716 (2006).
\bibitem{Seo06}  H.~Seo, K.~Tsutsui, M.~Ogata, and J.~Merino, J. Phys. Soc. Jpn. {\bf 75}, 
114707 (2006).
\bibitem{Udagawa07} M.~Udagawa and Y.~Motome, \prl {\bf 98}, 206405 (2007).

\bibitem{Shingai} M.~Shingai, S.~Nishimoto, and Y.~Ohta, unpublished.
\bibitem{Ohta94} Y.~Ohta, K.~Tsutsui, W.~Koshibae, and S.~Maekawa, \prb {\bf 50}, 13594 (1994).
\bibitem{Scalettar89} R.~T.~Scalettar, E.~Y.~Loh, J.~E.~Gubernatis, A.~Moreo, S.~R.~White, 
D.J.~Scalapino, R.~L.~Sugar, and E.~Dagotto, \prl {\bf 62}, 1407 (1989); A.~Moreo and 
D.~J.~Scalapino, \prl {\bf 66}, 946 (1991).
\bibitem{Moskvin03} A.~S.~Moskvin, J.~M\'alek, M.~Knupfer, R.~Neudert, J.~Fink, R.~Hayn, 
S.-L.~Drechsler, N.~Motoyama, H.~Eisaki, and S.~Uchida, \prl {\bf 91}, 037001 (2003).
\bibitem{Fye91} R.~M.~Fye, M.~J.~Martins, D.~J.~Scalapino, J.~Wagner, and W.~Hanke, 
\prb {\bf 44}, 6909 (1991)
\end{thebibliography}
\end{document}